\documentclass[sigconf]{acmart}
\copyrightyear{2021}
\acmYear{2021}
\acmConference[WSDM WebTour'21]{ACM WSDM Workshop on Web Tourism (WSDM Webtour'21}{March 12, 2021}{Jerusalem, Israel}
\acmBooktitle{ACM WSDM Workshop on Web Tourism (WSDM Webtour'21), March 12, 2021, Jerusalem, Israel}
\acmPrice{}
\acmDOI{}
\acmISBN{}
\usepackage{makecell}
\usepackage{graphicx}
\usepackage{subcaption}
\usepackage{natbib}
\usepackage{svg}  % Eis o pacote que queremos.

\graphicspath{ {./images/} }

\AtBeginDocument{%
  \providecommand\BibTeX{{%
    \normalfont B\kern-0.5em{\scshape i\kern-0.25em b}\kern-0.8em\TeX}}}

\begin{document}

\title{Hybrid Model with Time Modeling for Sequential Recommender Systems}


\author{Marlesson R. O. Santana}
\email{marlessonsa@gmail.com}
\affiliation{%
  \institution{Deep Learning Brazil -- Federal University of Goiás}
  \streetaddress{Av. Esperança, s/n - Chácaras de Recreio Samambaia}
  \city{Goiânia}
  \state{Goiás}
  \country{Brazil}
  \postcode{74690-900}
}

\author{Anderson Soares}
\email{anderson@inf.ufg.br}
\affiliation{%
  \institution{Deep Learning Brazil -- Federal University of Goiás}
  \streetaddress{Av. Esperança, s/n - Chácaras de Recreio Samambaia}
  \city{Goiânia}
  \state{Goiás}
  \country{Brazil}
  \postcode{74690-900}
}

\renewcommand{\shortauthors}{Santana and Soares.}

\begin{abstract}
Deep learning based methods have been used successfully in recommender system problems. Approaches using recurrent neural networks, transformers, and attention mechanisms are useful to model users' long- and short-term preferences in sequential interactions. To explore different session-based recommendation solutions, Booking.com recently organized the \textit{WSDM WebTour 2021 Challenge}, which aims to benchmark models to recommend the final city in a trip. This study presents our approach to this challenge. We conducted several experiments to test different state-of-the-art deep learning architectures for recommender systems. Further, we proposed some changes to Neural Attentive Recommendation Machine (NARM), adapted its architecture for the challenge objective, and implemented training approaches that can be used in any session-based model to improve accuracy. Our experimental result shows that the improved NARM outperforms all other state-of-the-art benchmark methods.

\end{abstract}





\maketitle

\section{Introduction}
Deep learning approaches for recommender systems have garnered significant attention owing to their potential for modeling long- and short-term user preferences.  Models using recurrent neural networks (RNNs), transformers, and attention mechanisms are handy for text, presenting encouraging results when used in session-based recommendation problems \cite{li2017neural, tan2016improved, volkovs2019robust, chen2019behavior, kang2018self, tang2018personalized}, where sequential information and short-term context are fundamental in recommending the next item of the session.

In general, session-based approaches use only the information from the item's interaction sequence in the session as a predictor of the next item. However, in some domains, contextual information within and between sessions and global information are essential to modeling user behavior. Recommending travel destinations has several particularities, such as the sequence of cities on a trip can contain noise, financial constraints may lead to diversions, and destinations are highly correlated to the moment in time and duration of the trip \cite{mizrachi2019combining, bernardi2019150, kiseleva2015go} .

The \textit{WSDM WebTour 2021 Challenge} \cite{booking2021challenge} organized by Booking.com, focuses on recommending travel destinations in a session. This study describes our approach for the \textit{WSDM WebTour 2021 Challenge} and proposes some approaches that can improve session-based models for recommender systems, especially with the highly time-dependent recommendation domains, noise information, and imbalanced classes. 

\section{The Challenge and Dataset}\label{sec:conclusions}

Booking.com is the world’s largest online travel agency. It is a platform where millions of travelers find accommodations for their trips, and millions of accommodation providers list their hotels, apartments, guest houses, and other lodgings. \cite{kiseleva2015go, bernardi2019150}. 

Booking.com recently organized the \textit{WSDM WebTour 2021 Challenge} \cite{booking2021challenge}. The training dataset consists of over a million anonymized hotel reservations based on real data. Each reservation is a part of a customer’s trip, which includes at least four consecutive reservations. The challenge's goal is to recommend the final city of each trip, and we evaluated models using an accuracy metric for the first four items suggested. For Accuracy@4, the metric value is one when the real city is one of the four main suggestions and zero otherwise.

% \begin{equation}
%   \lim_{n\rightarrow \infty}x=0
% \end{equation}
% Total: 1 166 835
% User Uniq: 200153
% city uniq: 39901
% country: 165
% trips: 217 686

% Date Time: 2015-12-31 - 2017-02-27

\section{Proposed Approaches}\label{sec:aproaches}

% This section describes each approach used in the present study to improve our final model and its architecture based on a state-of-the-art model. Each approach can be applied separately in other session-based models.

This section describes each approach used in the present study to improve our final model. First, we chose a state-of-the-art session-based model to improve (described in \ref{seq:architecture}). Further, we created new features from statistical information about users and cities and added time modeling to focus on the travel problem particularity (described in \ref{seq:feature_enginner} and \ref{sec:time_modeling}). Also, we have reduced the effect of the imbalance dataset using specific loss functions and multitask modeling (described in \ref{sec:loss_function} and \ref{sec:multi_task}). Finally, we improve the generalization of noise information through data augmentation (described in \ref{sec:data_arg}).

Each approach can be applied separately in other session-based models.

%  Further, we proposed some changes to Neural Attentive Recommendation Machine (NARM), adapted its architecture for the challenge objective, and implemented training approaches that can be used in any session-based model to improve accuracy

\subsection{Model Architecture}
\label{seq:architecture}

We use the architecture of the neural attentive recommendation machine (NARM) \cite{li2017neural} and adapt it for the traveling problem presented in this study. NARM uses an encoder-decoder architecture to address the session-based recommendations problems. According to the authors of the paper, the idea of NARM is to build a hidden representation of the current session using an RNN module. It converts the input click sequence $X = [x_1, x_2, ..., x_n]$ into a set of high-dimensional hidden representation with an attention signal that can be used to produce a ranking list of all items that can occur in the next step of the current session. 

Our approach uses categorical and dense features of the user, city, and trip combined with the trip history. The core of the NARM module is the same as the original paper, and we just changed the size of the inputs, the bottleneck with hidden representation, and the outputs.

The features are concatenated and follow two paths. The first group of features passes through an attention layer before being fed into the NARM module. It is an important step for relating different positions of the same input sequence. In the second path, the features bypass the feature bottleneck generated by the NARM module, which improves the decoder by providing it with more contextual information on the session.

\begin{figure}[h]
  \centering
  \includegraphics[width=250pt]{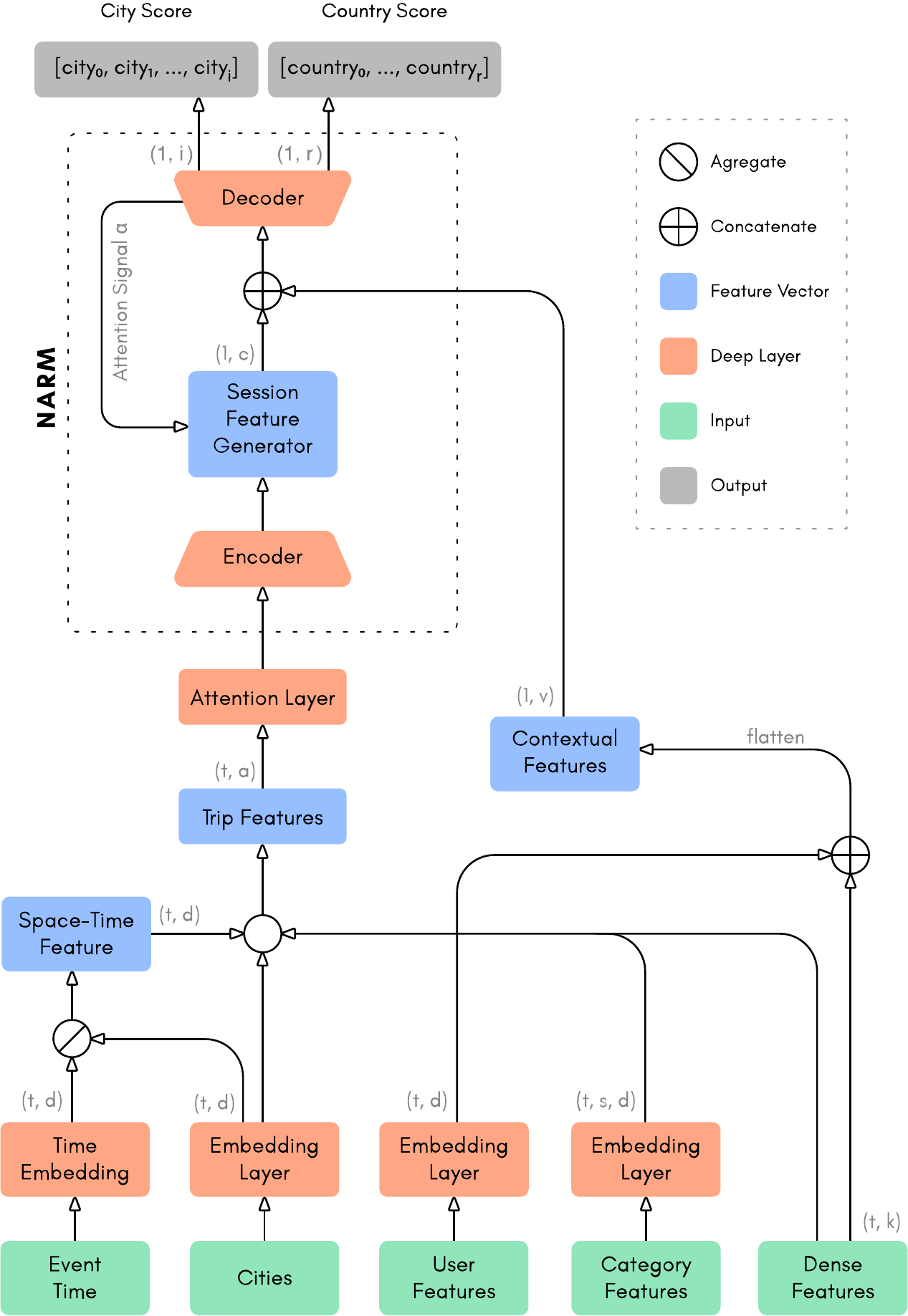}
%   \includesvg[width=250pt]{images/Diagrama_NARM.pdf}
  \caption{The general architecture of the proposed model. Our approach uses categorical and dense features of the user, city combined with the trip history. The features follow two paths. First, the features pass through an attention layer before being fed into the NARM module and in the second path, the features bypass the feature bottleneck generated by the NARM module, which improves the decoder by providing it with more contextual information on the session.}

%   \caption{The general architecture of the proposed model. We use categorical and dense features combined with the trip history for the NARM module.}
  \Description{The general architecture of the proposed model.}
  \label{fig:architecture}
\end{figure}

Figure \ref{fig:architecture} shows the final architecture proposed with modifications described in this study. 

% For more details about the NARM module, please consult the original publication \textit{"Neural Attentive Session-based Recommendation"} by \citeauthor{li2017neural}\cite{li2017neural}.

\subsection{Feature Engineer}
\label{seq:feature_enginner}

Our approach uses statistical features from users, cities, and trips combined with the trip's history information. For every trip, we extract more than 30 statistical features. Below we provide examples of the important features we use in the present study:

\textbf{User Statistics}
\begin{itemize}
    \item Number of trips
    \item Number of cities visited
    \item Average number of unique cities visited per trip
    \item Average trip size
    \item Average trip duration
    \item Most frequently traveled month
    \item Number of unique cities the user visited
    \item Number of unique countries the user visited
    \item Device used for booking
    \item Booker country
\end{itemize}

\textbf{City Statistics}
\begin{itemize}
  \item City interactions that have preceded the current trip
  \item City interactions that have preceded the current trip by user
  \item City interactions in the current trip
  \item Country interactions that have preceded the current trip
  \item Country interactions that have preceded the current trip by user
\end{itemize}

% \textbf{History Information}
% \begin{itemize}
%   \item List of cities in trip
%   \item List of Affiliate in trip
%   \item List of device used 
%   \item Checkin dates 
% \end{itemize}
Furthermore, we employ user statistical features to create user embedding from an autoencoder trained on the same training split data. This approach proved helpful because only 0.7\% of the users have made more than one trip; the vast majority are new users. In addition to providing a dense representation of the user features, the autoencoder ensures that similar users are close to each other in the created vector space. Moreover, the autoencoder was trained separately from the recommendation model. Figure \ref{fig:user_emb} shows user embeddings colored by the most frequently traveled month. 

\subsection{Time Modeling}\label{sec:time_modeling}

In terms of tourism, a trip's location is merely a piece of information that can tell how good the place is for a trip. The date (moment in time) and duration of the trip are as crucial as, for example, sometimes, the traveler can only visit certain places only in specific seasons or specific time windows. Sometimes, it is not feasible to stay for a few days more.

Embeddings are highly useful for modeling categorical features such as the location, and RNNs are remarkably good for capturing sequential data's time-dependence. However, they are limited when the events have irregular time intervals, such as duration of stay on a specific city trip. Therefore, we need to incorporate time interval explicitly into the model as a time-space embedding. In this case, we use the approach presented in \cite{li2017time} concatenated with a start trip month embedding for time-space embedding.

Therefore, each city embedding presented in a trip follows through a time embedding layer. We incorporate time (start trip month) and duration (how long users stayed in a city) information. In the end, we obtain a space-time embedding. 

\subsection{Multi-Task Learning}\label{sec:multi_task}

The data about the cities in our dataset are imbalanced and have high dimensionality. In contrast, the data about countries are less imbalanced and have lesser dimensionality. Thus, we adapt our model to predict both to try to improve models' gradient signal. According to \cite{ruder2017overview}, multi-task learning acts as a regularizer by introducing an inductive bias into the model. As such, it reduces the risk of overfitting and the Rademacher complexity of the model, and thus, it has the ability to fit random noise.

We used two targets in the present study. The city of the next step interaction was the main target that we used to evaluate the model, and the city's country was used as the second target. The second target was used only for regularization and as inductive bias in the primary target. Both targets use the same loss function, and the final loss is a combination of the two.

\subsection{Loss Function}\label{sec:loss_function}

Focal loss \cite{lin2017focal} is very useful for training imbalanced datasets. It adds a weighted term in front of the cross-entropy loss to balance the gradient from positive and negative samples. Easily classified negatives comprise most of the loss and dominate the gradient; this effect can be reduced by using a focal loss strategy. We use focal loss for both targets, and the final loss is the average of each loss.

Formally, the focal loss is expressed as follows:
\begin{equation}
  FL(p) = -\alpha_t(1-p)^\gamma log(p)
\end{equation}
where $\gamma$ is adjusts the rate at which easy examples are down-weighted and $\alpha$ is a prefixed value between 0 and 1 to balance the positive-labeled samples and negative-labeled samples. We use Equation \ref{eq:loss} for calculation loss. 
\begin{equation}\label{eq:loss}
  L(p_c, p_o) = \beta FL(p_c) + (1-\beta) FL(p_o)
\end{equation} where $\beta$ is a balance parameter, and $p_c$ and $p_o$ are obtained from the target prediction.

\subsection{Data Augmentation}\label{sec:data_arg}

Data augmentation techniques have been widely used to enhance image \cite{shorten2019survey}, text \cite{wei2019eda}, or recommendations based models \cite{tan2016improved,de2015artificial}. The principal inputs of the model in the challenge are cities in the current session's history. These inputs can have noise, some users may take a long or short trip, and users can jump one or more cities on a popular route. We applied three different data augmentations steps to improve the accuracy and generalize the model. 

The first is a sequence preprocessing step proposed in \cite{tan2016improved,de2015artificial}, where we generate new samples using each trip's time-step. Given an input training trip $[c_1, c_2, c_3, ..., c_n]$, we generate the sequences and corresponding labels $([c_1], c_2), ([c_1, c_2], c_3), ([c_1, c_2, ..., c_{n-1}], c_{n})$ for training. We filter only trips with more than four cities. 

For each sample in training, we randomly choose a step in the trip to change. Given an input training trip $[c_1, c_2, c_3, c_4]$, we change it in three ways: remove a step as a dropout layer and generate sequences such as $[c_1, c_2,  c_4]$;  replace with a mask token (unknown) and generate sequences with the same size but with mask $[c_1, c_2,  c_3, c_{UKN}]$ ; and replace with a similar city, such as $[c_1, s_2,  c_3, c_4]$ where $s_2$ and $c_2$ are similar cities. The mask token will be the same as that used in production/inference mode when the model does not embed a specific city, and the similarity definition is the same as that we used for the Item-KNN model. 

We apply both methods to make the model less susceptible to overfitting or noise information. 

\section{Experiments}\label{sec:experiments}
In this section, we present the experimental settings and results.

\subsection{Dataset}

We randomly partition the training dataset by trip, using 90\% of data for training and 10\% for model validation. We filtered trips with less than four cities or more than 10 and trips with a duration of more than 22 days in the training split. These values came from the initial analysis and are outliers. Table \ref{tab:data_statistics} shows the statistics of the dataset used in our experiments.

\begin{table}[!h]
  \caption{Statistics of the dataset used in our experiments}
  \label{tab:data_statistics}
  \begin{tabular}{ccccl}
    \toprule
    Split & Trips & Users & Cities\\
    \midrule
    train &  195917 & 181480 & 38542 \\
    test  &  21769 & 21524 & 15488 \\
    % valid & 378667 & 70662 & 68502 & 21305 \\
  \bottomrule
\end{tabular}
\end{table}

\subsection{Baselines}

To show the effectiveness of our approach, we choose some popular baseline models used for session-based recommendation problems.

\begin{itemize}
  \item \textbf{Popularity}: Popular predictor recommends the most popular city in last city's country on the current trip.
  \item \textbf{Item-KNN}: This baseline recommends the most similar city to the last city on current trip. The similarity is defined as the cosine distance between the trip vector of the city.  It is similar to the co-occurrence approach.  \cite{davidson2010youtube}
  \item \textbf{Caser}: This baseline is a convolutional neural network (CNN) approach for a sequential recommendation. The Convolutional Sequence Embedding Recommendation Model (Caser) embeds a sequence of recent items into an "image" in time and learns sequential patterns as local features. \cite{tang2018personalized}
%   \item \textbf{Transformer}: The transformers approach is popular in natural language processing (NLP) problems today as it understands language better than the RNN models. We used only the encoder module from the transformer architecture, which is a similar approach to that of Layer 6 AI \cite{volkovs2019robust} at the 2019 in Recsys Challenge. 
  \item \textbf{SASRec}: SASRec is a self-attentive approach that captures users’ sequential behaviors and achieves state-of-the-art performance on sequential recommendation. \cite{kang2018self}
  \item \textbf{NARM}: Neural Attentive Session-based Recommendation is an encoder-decoder architecture with an attention mechanism to model the user's sequential behavior, which is then combined as a unified session representation later. \cite{li2017neural}
\end{itemize}

We choose the NARM model, which shows the best performance among the models listed above, such as our baseline model to improve during the competition. Therefore, our approach is an adaptation of the NARM model. Some approaches were modeled for the click prediction problem. For our experiments, we adapted the last layer of all deep learning models to produce a ranking list of all items,  $ y = [ y_{1}, y_{2}, y_{3}, ..., y_{n} ] $, that can occur on the current trip. Furthermore, we trained our model using the same loss function.

\subsection{Experimental Settings}

All our models were trained using similar parameters. We used 50-dimensional embeddings for all category features, and we normalized dense features using standard deviation. Optimization was carried out using Rectified Adam (RAdam), with a 0.001 learning rating and 0.01 weight decay, with mini-batch size fixed at 64. We truncated the trip history size using a fixed window of 10 time-steps with padding. The number of epochs was defined by early stopping using 10 steps, and we used cross-entropy as a loss function for most models or focal loss with 1 $\alpha$ and 3 $\gamma$. 

Finally, we use the MARS-Gym framework \cite{santana2020marsgym} to model, train, and evaluate all experiments described in this paper. All experiments are present and can be reviewed at \url{https://github.com/marlesson/booking_challenge}.

\subsection{Experimental Results}

\begin{figure}[h]
  \centering
  \includegraphics[width=230pt]{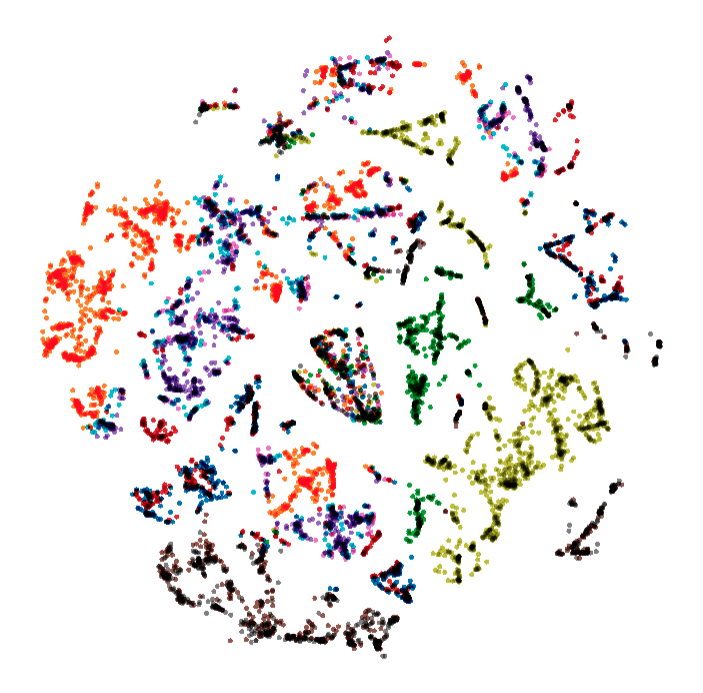}
  \caption{We used t-SNE to project 100-D into 2-D and produce this Figure. Each point represents a user, and the color shows the most frequent month that users traveling.}
  \Description{User embedding}
  \label{fig:user_emb}
\end{figure}

Each user's embeddings are presented in the 2-D plane in Figure \ref{fig:user_emb}. We can see that the month is a good predictor of user behavior, and, in general, users who travel in the same month also share similarities. We can also notice that there is a mixed cluster in the center, maybe belonging to users who are likely to travel on different dates with greater frequency. Furthermore, we can use this representation for new users to improve a cold-start recommendation.

Each model's performance on the test dataset is summarized in Table \ref{tab:freq}. We can see that the Item-KNN method performs similarly to the Popularity method, which is a cold start method. Therefore, there is no benefit in adopting the Item-KNN method given the complexity posed by cold recommendations. 

% The same can be said for the Transformer method, which, although producing better results compared with the Item-KNN method, suffers from high complexity. However, this is surprising considering attention-based approaches such as SASRec and NARM performed well using the same data.

% \begin{table}[h!]
%   \caption{Performance comparison of proposed method with baseline methods}
%   \label{tab:freq}
%   \begin{tabular}{lccc}
%     \toprule
%     Methods & ACC@4 (test) & ACC@4 (final)\\
%     \midrule
%     Popularity & 0.364 & -  \\
%     Item-KNN & 0.371 & -  \\
%     Transformer & 0.407 & -  \\
%     SASRec & 0.468 & -  \\
%     Caser & 0.477 & -  \\
%     NARM & 0.492 & -  \\
%     NARM V1 & 0.520 & -  \\
%     NARM V2 & 0.543 & -  \\
%     \bottomrule
%     Improved NARM (ensamble) & 0.574 & -  \\
%     \bottomrule

% \end{tabular}
% \end{table}

\begin{table}[h!]
  \caption{Performance comparison of proposed method with baseline methods}
  \label{tab:freq}
  \begin{tabular}{lccc}
    \toprule
    Methods & ACC@4 (test)\\
    \midrule
    Popularity & 0.364   \\
    Item-KNN & 0.371   \\
    % Transformer & 0.39   \\
    SASRec & 0.478   \\
    Caser & 0.484   \\
    NARM & 0.497   \\
    NARM V1 & 0.520   \\
    NARM V2 & 0.545   \\
    % \bottomrule
    % NARM V2 (ensemble) & 0.569  \\
    % \bottomrule

\end{tabular}
\end{table}

SASRec and Caser are very different approaches, the former based on attention mechanism and later on horizontal and vertical convolutions. Both showed similar results, with Caser performing slightly better. However, the best base baseline model was NARM. We can see that the original NARM outperformed state-of-the-art baselines using the same input information. Therefore, we chose NARM in this study to improve the architecture and training phase. 

We evaluate two versions of NARM. In NARM V1, we use only improvements  that were applied in the training and regularization steps, with the same input data were used for other baseline models, but applying approaches \ref{sec:time_modeling}, \ref{sec:multi_task}, \ref{sec:loss_function}, \ref{sec:data_arg}. NARM V1 outperforms the original NARM with approximately $+4.62\%$ accuracy. This is a promising finding, as this indicates that these techniques can now be used for any other session-based models that are in need of improvement.

Finally, we apply all approaches present in this study to NARM V2, and it represents our final approach for the challenge. Figure \ref{fig:architecture} shows our final architecture. We found that NARM V2 outperforms the original NARM with approximately $+9.66\%$ accuracy. This supports the idea that when it is not possible to obtain an item or user metadata, session statistics can also be used as features.

\section{Conclusion}

In this study, we present our approach for the \textit{WSDM WebTour 2021 Challenge}. We conducted several experiments using different session-based models to recommend the next destination on a trip. We modified the existing NARM model to add contextual information to the session, space-time modeling, and approaches to reduce the negative effect of class imbalance in the training phase. Our results show that it is possible to enhance the performance of state-of-the-art models through simple changes. The improved NARM outperforms all baseline models, and that the implemented training approaches can be used in any session-based recommendations model.

% \input{tex/acknowledgments}

% \bibliographystyle{ACM-Reference-Format}
% \bibliography{main}

\bibliographystyle{ACM-Reference-Format}
\bibliography{main}

% \newpage
% \appendix
% \input{tex/appendices}

\end{document}